# Predicting synthesizable cobalt and manganese silicides, germanide with desirable magnetic anisotropy energy

Ze-Jin Yang
School of Science, Zhejiang University of Technology, Hangzhou, 310023, China
zejinyang@zjut.edu.cn

**Abstract**

The nanoparticle $Co_3Si$ ($P6_3/mmc$) displays remarkable magnetism [Appl. Phys. Lett. 108, 152406 (2016)], we thus searched cobalt silicides and several phases are searched including a *Cmcm* with 60 meV/atom lower than that of $P6_3/mmc$. A lower-energy Co $R\bar{3}m$ (-7.03 eV/atom) is predicted, whose energy is higher than that of known $P6_3/mmc$ (-7.04 eV/atom) but is lower than that of $Fm\bar{3}m$ (-7.02 eV/atom). Three small-magnetism low-energy $Fe_5Si_3$ structures are searched with energies 30 meV/atom lower than that of experimental $P6_3/mcm$. The strong lattice shape dependence of magnetocrystalline anisotropy energy (MAE) is studied through $X_5Si_3$ (X=Mn, Fe, Co). The building-block shape and energy order of cobalt silicide is dominated by Co $P6_3/mmc$, $Fm\bar{3}m$ , $R\bar{3}m$ , respectively. The $Co_3C$ and $Co_3Sn$ have positive formation of energy, thus only $Co_3Ge$ has similar structures with those of counterparts of $Co_3Si$. Several low-energy perfect or nearly-perfect easy-axis/plane MAE $Mn_3Si$, $Mn_5Si_2$, and $Mn_5Si_3$ structures are searched and present important application, as is also the case in Ge-containing counterparts. A structure $I4_122$ with energy 300 meV/atom lower than that of experimental $Mn_5Si_3$ $P6_3/mcm$ is searched.

## I. Introduction

Magnetic materials play an important role in our life, such as hard disk in laptop. Magnetism comprises of diverse research fields in physics, for instance giant magnetoresistance effect, colossal magnetoresistance effect, magnetocaloric effect, magnetic exchange bias effect, magnetic proximity effect, spintronics, and spin Seebeck effect. Recently, the most frequently used permanent magnets are $Nd_2Fe_{14}B$ and $SmCo_5$-based compounds, both containing rare-earth element. Due to limited rare-earth resources, therefore it is necessary to synthesize materials without rare-earth element[1]. Magnetocrystalline anisotropy energy (MAE) plays a key role in the development of magnetic storage media. Different MAE magnitudes have extensive applications in the fields of soft and hard magnetic materials.

Nanoscale $Co_3Si$ (*P*$6_3$*/mmc*) is synthesized[2] and displayed excellent easy-plane MAE in its nanoparticle form, based on which the authors conclude that a path towards fabrication of nanoparticle materials with easy-plane anisotropies, valuable energy products, and in particular containing earth-rich substance is successfully built. In fact, metastable hexagonal $Co_3Si$ (*P*$6_3$*/mmc*) is observed long time ago[3-5]. However, the discrepancy between their theoretical calculation for the ideal $Co_3Si$ nanoparticale (-64 Merg/$cm^3$) and experimental estimation (48 Merg/$cm^3$) anisotropy is large, possibly due to the experimental surplus of the 10 vol. % Co in the nanoparticle composite. Unexpectedly, the extra Co fails to form the other stoichiometry compounds such as $Co_4Si$, *etc*. Considering such important conclusion, it is very necessary to further study the cobalt silicides carefully. Moreover, nonequilibrium fabrication technique usually synthesizes metastable phases with very large MAE. Therefore, we use the advanced crystal structure prediction method to search the possible low-energy structure. Moreover, the observed $Co_4Si$ also is a metastable phase but its structure is still unknown[4]. Three different phases are observed in $Co_2Si$ till now, it is $\alpha$-$Co_2Si$, $\beta$-$Co_2Si$, $\gamma$-$Co_2Si$, respectively[3-5]. These phenomena suggest that there are many metastable phases in these low Co:Si stoichiometry ratio compounds.

The same group[2] also observed that nanoparticles $Mn_5Si_3$ presents appreciable magnetocrystalline anisotropy constants ($K_1$=6.2 Mergs/cm$^3$ at 300K and at 12.8 Mergs/cm$^3$ at 3K)[6]. Mn-based structures became an important study focus from the viewpoints of designing rare-earth-free materials with useful spin-related applications[7]. Thus, we carefully searched the experimentally known but simple Mn-based compounds, such as $Mn_3Si$[8], $Mn_5Si_2$[9], $Mn_5Si_3$[8], $Mn_6Si$[10], $Mn_9Si_2$[11]. We didn't search the other stoichiometry because the search and MAE calculation are extremely time-consuming. Considering the important stoichiometry of 5:3, we also searched the $Fe_5Si_3$ as the previous search might miss[1] some energetically similar structures. Experimental structure has 53 atoms per unit cell ($Mn_{85.5}Si_{14.5}$)[10] in $Mn_6Si$, but it is 56 atoms for the 8-unit cell, the same authors also found 186 atoms for ($Mn_{81.5}Si_{18.5}$)[11] in $Mn_9Si_2$, whereas it is 176 atoms for 16-unit cell. Thus it is difficult to compare the searched perfect structures with their data. In this study, we thoroughly searched the binary magnetic cobalt and manganese silicides as well as $Fe_5Si_3$ in order to find the ideal MAE, fortunately, our searched many structures have very interesting properties in the diverse magnetic field.

## II. Methods

We searched the structures by CALYPSO (Crystal structure AnaLYsis by Particle Swarm Optimization) code[12, 13]. The obtained structures are *ab initio* optimized[14, 15] by projector-augmented wave method[16]. The spin polarized generalized gradient approximation and the Perdew-Burke-Ernzerhof functional are used to model the exchange-correlation energy[17]. The structure is optimized with the electronic convergence below $10^{-7}$ eV, the k-point density is below $2\pi\times0.02$ Å$^{-1}$ in the Brillouin-zone. The force on the atom is converged to less than 0.01 eV/Å during the relaxation. Energy cutoff for the plan-wave basis is 350 eV. These parameters are sufficient to ensure the reliability of the calculated results based on our calculations on iron silicides[1]. The on-site coulomb interaction term $U$ (and $J$) on the energy is neglected during all the structure computation[14, 15] based on the previous conclusion of iron silicides[1].

## III. Results and discussion

### A. MAE of experimental nanoparticle size $Co_3Si$

We define the lowest-energy axis as the reference axis in this paper. Experimentally[2] synthesized nanoscale $Co_3Si$ ($P6_3/mmc$) with an easy-plane anisotropy with a high $K_1$=-64 Merg/cm$^3$ (-6.4 MJ/m$^3$ ) and magnetic polarization of 9.2 kG (0.92 T), which is confirmed by their calculated 64-atom $Co_3Si$ nanoparticle. In addition, they also found that $Mn_5Si_3$ nanoparticle also presents excellent properties[6]. Our calculated $Co_3Si$ shows $a$=$b$=4.9794 Å, $c$=3.9755 Å, substantially different with the experimental data $a$=$b$=4.99 Å, $c$=4.497 Å. Our calculated absolute total-energy difference between spin orientations in the (100) and (001) directions yields a MAE of 2.0891 meV per unit cell (containing 2 formula units or 8 atoms) for the bulk $Co_3Si$ crystal, which corresponds to $|K_1|$=39.21 Merg/cm$^3$ (or 261.1412 $\mu$eV/atom), far smaller than the experimentally reported value -3.9 meV ($K_1$=-64 Merg/cm$^3$), whose absolute magnitude is large and comparable to that of typical rare-earth easy-axis intermetallics, as said in that paper. Another $P6_3/mmc$ ($Co_3Si$), with an average energy of -6.8931 eV/atom, shows shortened lattice $b$ and slightly-changed $a$ and $c$, as a result which induces a nearly-perfect easy-plane MAE, with $E_{001}$=80.28 and $E_{100}$=0.06 $\mu$eV/atom (or energies along $c$ and $a$ axes), respectively and similarly hereinafter. Thus, the perfect bulk crystal failed to obtain the corresponding magnetism of nanoparticle due probably to the surface effect.

To reveal the possible origin of the giant axial energy difference, the nanoparticle is constructed by screening the neighbor interaction through the long vacuum layer larger than 10 Å. Table S1 shows that the perfect 8-atom, 32-atom, 64-atom nanoparticle could produce MAE with values about 0.1, 0.07, 0.12 meV/atom, respectively, which still couldn't repeat the experimental MAE (-3.9 meV per unit cell or -0.4875 meV/atom). We didn't calculate the case of 128-atom unit cell because its atomic arrangement is totally same with the cases of 8/32/64-atom. One possible explanation is that the experimentally synthesized structure exists large distortion along the lattice parameter $c$ orientation. Distortion usually induces large MAE but it is probably meaningless in the practical industrial application. More importantly, distortion usually couldn't produce perfect MAE. In addition, their lattice parameter *c* is also couldn't be explained from the unknown surface substance or internal defect structure. The surface interaction is almost impossible to elongate the lattice parameter *c* with such large value, from about 4.0 to 4.5 Å, thus we ignore to calculate the MAE at the case of *c*=4.5 Å because such structure is hypothetical by uniaxial strain and is also unstable.

The experimental $Co_3Si$ *P*$6_3$*/mmc* has a ~10% elongation of lattice *c* which induces the giant distortion of lattice shape and break the atomic equilibrium status. We test it by the $X_5Si_3$ (X=Mn, Fe, Co), as is shown in Table **S2**, in which the strong lattice framework or atomic occupancy dependence is observed, the average energy difference between easy and hard axes of the three lattice has intrinsic value, such as the stable $Co_5Si_3$ has an MAE value of ~13 $\mu$eV/atom, substitutions of Co by Mn and Fe produce ~29 ($Co_5Si_3$) and ~23 ($Fe_5Si_3$), respectively. For the case of $Mn_5Si_3$, those corresponding values are ~39 and ~39 ($Co_5Si_3$) and ~56 ($Fe_5Si_3$). For the case of $Fe_5Si_3$, they are ~157 and ~88 ($Mn_5Si_3$) and 94 ($Co_5Si_3$), respectively.

**B. The searched lower-energy structures than that of $Co_3Si$ *P*$6_3$*/mmc***

We only searched the stoichiometry of Co:Si>1 in order to find the structures with large magnetism. The current search obtained 5 lower energy structures than that of experimentally synthesized $Co_3Si$ *P*$6_3$*/mmc* (-6.9 eV/atom), in which *Cmcm* structure is about 60 meV/atom lower than that of *P*$6_3$*/mmc*. The other lower-energy phases include *Pmmn*, (-6.9143), *I*4*/mmm* (-6.9197), *I*4*/mcm* (-6.9025), *P*$2_1$ (-6.9071 eV/atom), as is shown in Table **1**. In fact, there are also several phases have only slightly higher energies than that of *P*$6_3$*/mmc*. In other words, the metastable phase with energy about several dozens of meV/atom lower than that of experimental phase or the searched lowest-energy phase (probably be) still could be synthesized.

The currently calculated MAE of $Co_3Si$ *P*$6_3$*/mmc* is 261.14 $\mu$eV/atom with an easy-axis energy distribution, despite which is far smaller than that of previously calculated 487.5 and measured 365.625 $\mu$eV/atom but ~260 is still a very large value. Thus the MAE of $Co_3Si$ *P*$6_3$*/mmc* (261.14 $\mu$eV/atom) is not comparable but is far smaller than that of typical rare-earth easy-axis intermetallics compounds.

$Co_3Si$ *P*$6_3$*/mmc* ($a$=$b$=4.9794, $c$=3.9754 Å) is the same lattice framework with that of Co *P*$6_3$*/mmc* ($a$=$b$=2.507, $c$=4.069 Å), that is, the 2×2×1 supercell of Co is $a$=$b$=5.014 Å, thus $Co_3Si$ and Co have totally same lattice with the only exception of the slight distortion due to the substitution of Co by Si. Projections of the $Co_3Si$ *P*$6_3$*/mmc* towards *a* and *b* axes are totally same owing to the inversion symmetry operation nature, thus the MAE along *a* and *b* axes are identical. Moreover, the mixture of nanoparticle $Co_3Si$ with about 10 vol. % $\varepsilon$-Co makes the resolution of this question very difficult. The building block of *P*$6_3$*/mmc* is tetrahedron with the distance of Co~Si=2.4454 Å and Co~Co=2.4671 Å. The building block of *Cmcm* is a cubic body but the Si doesn't distribute at the cubic center, thus presenting two different kinds of distance, it is 2.4455 and 2.4446 Å in its first shell or nearest neighbor, namely, a slightly distorted bulk centered cubic (BCC) unit.

*Cmcm*-A (named as A … because many structures have same symmetry but show obviously different energies) is the searched lowest-energy phase with $E_{010}$=0.24 and $E_{100}$=28.46 $\mu$eV/atom, respectively. Its building block shape is most similar with that of $P6_3/mmc$. In detail, the two neighboring layer "$Co_3Si$" has a torsion angle 60° in $P6_3/mmc$ but it is zero degree in *Cmcm*-A. Moreover, the general shape of building block in *Cmcm*-A is more similar with a BCC, the distance of $Co_9$~$Co_3$=$Co_9$~$Co_{14}$=2.7097 Å, $Co_{11}$~$Co_{14}$=$Co_{11}$~$Co_3$=2.547 Å. The distance between the central Si and its 8 first-nearest neighbors are $Co_{2,3}$~$Si_1$=$Co_{13,14}$~$Si_1$=2.3227 Å, $Co_{5,9}$~$Si_1$=2.3796 Å, $Co_{7,11}$~$Si_1$=2.5162 Å. Its $Co_1$~$Si_2$=$Si_1$~$Co_8$=4.1856 Å corresponds to the 4.2992 and 4.3254 Å in $P6_3/mmc$. Thus the phase transition from $P6_3/mmc$ to *Cmcm*-A in $Co_3Si$ is possible under high pressure. The central Si deviates the geometric position evidently in *Cmcm*-B relative to that of *Cmcm*-A, for instance, the four pairs Co~Si distances are 2.3837 Å in first shell but the other four ones are 2.482 Å in *Cmcm*-B. Meanwhile, for the case of the second shell or second nearest neighbors, the Co~Si distances at the short first-shell side (2.3837 Å) is also short (2.3372 Å), vice versa, the other side (2.482 Å) corresponds to 2.8817 Å, the other pair Co~Si distances within the same horizontal plane show small variation, 3.1959 versus 3.1963 Å, only the two totally-symmetric Co present identical Co~Si distance, such as the distance between the top and bottom of the building block is 2.5666 Å. However, *Cmcm*-B ($\beta$=90.0002°) and *Cmcm*-A ($\beta$=90°) have large MAE difference due probably to the distorted angle. *Cmcm*-C is a higher-energy layered structure, whose two Si layers are separated by three Co layers, projections towards *a* and *c* axes show similar layer separation feature. Its building block is not tetragonal shape but triangle nested shape. Co~Si distances, whether in the same plane or the nearest neighbors, are not exactly an isosceles triangle. The $\beta$ is 90.0051 in *Cmcm*-D, meaning an even larger distortion, whose building block is more closer with that of Co $Fm\bar{3}m$, as a result its energy is higher than the structures packed by Co $P6_3/mmc$.

The searched $Co_3Si$ *I4/mmm*-A, (-6.9197 eV/atom) is about 19 meV/atom lower than that of $P6_3/mmc$ and has a perfect easy-plane MAE, with a value of $E_{010}$=141.42 $\mu$eV/atom along the hard *b* axis, which is evidenced by the identical projections towards *a* and *c* axes and the lattice parameters *a*=*c*=5.0558, *b*=6.6642 Å. Considering the multiplicity of the lattice shape we only calculate the axial energy along three lattice parameters as this is the most extensively used directions.

The searched $Co_3Si$ *I4/mmm*-B (-6.9181 eV/atom) is about 18 meV/atom lower than that of $P6_3/mmc$ and also shows totally-perfect easy-plane MAE with a value of $E_{100}$=221.61 $\mu$eV/atom along the hard *a* axis. Its perfect MAE could be attributed partially to their totally-symmetric arrangements at the both sides of the mirror that is formed by the *c*-axis parallel line along the angle bisector between *a* and *b* axis. In fact, we obtain several structures with same symmetry *I4/mmm* but their MAE are evidently different owing to the slight strain, which will cause large axial-energy difference, such as $E_{001}$=153.08 and $E_{010}$=25.22 in *I4/mmm*-C but $E_{001}$=0.03 and $E_{010}$=148.42 $\mu$eV/atom in *I4/mmm*-D, as is shown in Table 1. For simplicity, we ignore to compare their bond distances as they are generally similar with the cases of *Cmcm*. The other structures also have *I4/mmm* symmetry and large magnetism but present non-perfect MAE, indicting the complicated MAE feature.

The building block of $Co_3Si$ *Pmmn* (-6.9143 eV/atom) is similar with that of Co *P*$6_3$*/mmc*, whose MAE differs each other along the three axes and shows large discrepancy.

The *I*4*/mcm* (-6.9025 eV/atom) has nearly same energy with that of *P*$6_3$*/mmc*, whose $E_{001}$=2.8 and $E_{100}$=211.67 $\mu$eV/atom, respectively. This is also a structure with the MAE larger than 200 $\mu$eV/atom and shows nearly-perfect easy-plane anisotropic distributions. Despite the building block has large distortion relative to the perfect BCC, the inter-atomic distance still has certain symmetry, for example, $Co_{1,6,7,10}$-$Si_1$=2.4223, $Si_1$-$Si_{2,3}$=3.6626, $Co_{2,5,9,12}$-$Si_1$=2.7471, $Co_{3,4,8,11}$-$Si_1$=2.2918 Å, respectively. That is, the average distance between $Co_{2,3,8,9}$~$Si_1$ and $Co_{4,5,11,12}$~$Si_1$ are identical in the first shell, which contributes to the low energy of the lattice. Moreover, the second-shell atom also shows highly symmetric arrangements around the $Si_1$, which might be the origin of the nearly-perfect MAE. In a word, the MAE indeed depends on the atomic arrangements.

*P*$2_1$ (-6.9071 eV/atom) shows a distorted first and second shells and thus presents 8/6 different Co-Si distances in the first/second shells, respectively, whose first-shell distance could generally be divided into two different sorts with values of about 2.38 and 2.48 Å, respectively, whereas the second-shell distance presents less similarities, including $Co_1$-Si=2.5398, $Co_{13}$-Si=2.4813, $Co_6$-Si=3.048, $Co_8$-Si=3.22, $Co_7$-Si=3.1413, $Co_9$-Si=2.2983 Å, respectively. All of the neighbors of the Si are surrounded by the Co atoms, which might contributes to its lower energy despite the large distortion. Its axial energies are $E_{001}$=81.08 and $E_{100}$=137.16 $\mu$eV/atom, respectively, in accordance with the case of the largely-disordered atomic arrangements.

$Pm\bar{3}m$-A (-6.8891 eV/atom) presents nearly-perfect easy-axis MAE with values of $E_{010}$=71.9 and $E_{100}$=72.1 $\mu$eV/atom, respectively, whereas $Pm\bar{3}m$-B (-6.8876 eV/atom) shows perfect easy-plane MAE with a value of $E_{100}$=32.14 $\mu$eV/atom. *P*4*/mbm*-A (-6.8976 eV/atom) also shows nearly-perfect easy-plane MAE with values of $E_{001}$=2.14 and $E_{100}$=144.51 $\mu$eV/atom, respectively. The other searched $Co_3Si$ is shown in Table S3, in which $Fm\bar{3}m$-A (-6.8836 eV/atom) shows giant MAE with respective values of $E_{001}$=375.13 and $E_{010}$=877.91 $\mu$eV/atom. Most of the searched low-energy $Co_3Si$ has the similar building blocks with that of stable Co $P6_3/mmc$, whereas the majority of the high-energy $Co_3Si$ has the similar building blocks with that of metastable Co $Fm\bar{3}m$, namely, the building-block shape of the Co dominates the atomic arrangement and energy order of $Co_3Si$. Since $Co_3Si$ could present diverse structures with various magnetism we therefore also searched $Co_3C$, $Co_3Ge$, $Co_3Sn$, respectively. Unfortunately, only $Co_3Ge$ has negative formation of energy.

**C. The searched higher-energy structures of $Co_3Si$ than that of $P6_3/mmc$**

The following structures are several meV/atom higher than that of experimental $P6_3/mmc$ and might also be synthesized. A same-symmetry structure $P6_3/mmc$-B ($E_f$=-6.8931 eV/atom) with nearly-perfect easy-plane MAE is searched, whose $E_{001}$=80.28 and $E_{100}$=0.06 $\mu$eV/atom, respectively. $Pm\bar{3}m$-A ($E_f$=-6.8891 eV/atom) presents nearly-perfect easy-axis MAE with energies of $E_{010}$=71.9 and $E_{100}$=72.1 $\mu$eV/atom. $Pm\bar{3}m$ -B shows perfect easy-plane MAE ($E_f$=-6.8876 eV/atom) with an energy of $E_{100}$=32.14 $\mu$eV/atom. *P4/mbm*-A still presents nearly-perfect easy-plane MAE ($E_f$=-6.8976 eV/atom) with energies of $E_{001}$=2.14 and $E_{100}$=144.51 $\mu$eV/atom, respectively. These rich MAE could satisfy the diverse industry requirements.

**D. Predicted low-energy structure of $Co_3Ge$**

The symmetry of the searched low-energy $Co_3Ge$ structure is generally same with that of $Co_3Si$, as is shown in Table **2** and **S4**. The searched two lowest-energy structures also are *Cmcm* (-6.5163) and *I4/mmm* (-6.4817 eV/atom), with an energy difference of 35 meV/atom. The general bond length (or side length) of the most stable *Cmcm* structure is about 2.59 and 2.48 Å in the unit-formula parallelogram, and the longer Co-Ge distance is about 4.4 Å, as is plotted by the red line. We ignore to plot the building block as it is similar with that of $Co_3Si$. The $E_{010}$=0.24 and $E_{100}$=28.46 $\mu$eV/atom in $Co_3Si$ *Cmcm* corresponds to $E_{010}$=14.91 and $E_{100}$=32.19 in $Co_3Ge$. Similarly, a perfect and large easy-axis MAE ($E_{001}$=$E_{010}$=261.14 $\mu$eV/atom) in $Co_3Si$ (*P*6<sub></sub>3*/mmc*) corresponds to $E_{001}$=332.58 and $E_{010}$=85.53 $\mu$eV/atom in $Co_3Ge$ ($P6_3/mmc$). $P6_3/mmc$ has similar atomic arrangement with that of $P\bar{1}$ in $Co_3Ge$ but the latter presents nearly-perfect easy-axis MAE with values of $E_{010}$=116.38 and $E_{100}$=114 $\mu$eV/atom, and the energy of $P\bar{1}$ is lower than that of $P6_3/mmc$, with a difference of about 15 meV/atom, still revealing the strong atomic coordinate dependence of MAE.

Two $Co_3Ge$ *I4/mmm* structures have perfect easy-plane MAE, with values of $E_f$=-0.0795 eV/atom and $E_{001}$=157.06 $\mu$eV/atom, $E_f$=-0.0524 eV/atom and $E_{010}$=9.6218 $\mu$eV/atom, respectively. In fact, many *I4/mmm* structures show nearly-perfect easy-axis/plane MAE, as is shown in Table **S4**, such as one structure shows $E_{001}$=156.9 and $E_{010}$=156.95 $\mu$eV/atom, respectvely, whereas all of these *I4/mmm* structures have nearly identical energies, which might be a great challenge during experimental synthesization.

*Pmmm* ($E_f$=-0.0728 eV/atom) has nearly-perfect easy-axis MAE with values of $E_{001}$=122.41 and $E_{100}$=122.58 $\mu$eV/atom, respectively. The other same-symmetry structures (*Pmmm*) are shown in Table **S4**, which further demonstrates that the MAE is strongly dependent on the atomic coordinate. Similarly, the following structures have energies about 70 meV/atom higher than that of lowest-energy structure *Cmcm* and also might be synthesized, such as *Cmmm* presents medium MAE with values of $E_{010}$=25.5 and $E_{100}$=26.3, $P2_12_12_1$ has $E_{010}$=24.15 and $E_{100}$=23.25, *C*2/*m* has $E_{001}$=26.18 and $E_{100}$=23 $\mu$eV/atom, and so on, these nearly-perfect easy-axis MAE structures have potential applications in the low-magnetism requirement field. Several structures show nearly-perfect easy-plane MAE such as *Cm* presents $E_{001}$=40.92 and $E_{100}$=2.64, *Ama*2 presents $E_{001}$=269.05 and $E_{100}$=0.78, $C222_1$ presents $E_{001}$=112.07, $E_{010}$=4.75 $\mu$eV/atom, and so on. *C*2/*m* shows relatively large MAE with values of $E_{010}$=42.56 and $E_{100}$=108.01 $\mu$eV/atom, respectively, all of the above structures might be synthesized. In a word, both $Co_3Ge$ and $Co_3Si$ could crystalline many low-energy structures with diverse coercivity magnetism. These low-energy $Co_3Ge$ structures are also dominated by the building block of Co $P6_3/mmc$, as is plotted by the red line.

**E. Predicted Co structures**

To determine the common building block shapes, we further searched the Co structure. Many low-energy Co structures are searched, as is shown in Table 2, in which one new phase is predicted, $R\bar{3}m$ (-7.0306 eV/atom), whose energy is well located at the intermediate range of the two experimental phases $Fm\bar{3}m$ (-7.0206) and $P6_3/mmc$ (-7.0398 eV/atom). Two new phases including $P4_2/mnm$ (-6.997) and $Cmca$ (-6.9972 eV/atom) are searched with energies of 25 meV/atom higher than that of $Fm\bar{3}m$. The magnetic moment are 1.623, 1.669, 1.65 $\mu_B$ for $P6_3/mmc$, $Fm\bar{3}m$, and $R\bar{3}m$, respectively. All of these structures have similar building blocks with the exception of certain distortion. Si and C have complicated structures and also have been searched for many times by the researcher in recent years, thus it is not necessary to search them again.

**F. Predicted cobalt silicides excluding $Co_3Si$**

Figure 1 shows the $E_f$ of the binary cobalt silicides. For simplicity, we connect the lowest-energy structures directly because whether some of the searched lowest-energy structures could be synthesized or not depends on many factors, such as its own energy, the synthesized techniques, the decomposed products, and so on. As is shown in Table S5, CoSi crystallines in $P2_13$ at ambient conditions, we searched one metastable phase $P2_1/c$, it is about 30 meV/atom higher than that of $P2_13$, unfortunately it still presents zero magnetism. $Co_2Si$ stabilizes[18] in $Pnma$ ($E_f$=-0.4501 eV/atom) and shows $E_{001}$=32.03 and $E_{010}$=18.96 $\mu$eV/atom, respectively. We predicted one metastable $P2/m$ ($E_f$=-0.4011 eV/atom) with nearly-perfect easy-axis energies of $E_{010}$=23.45 and $E_{100}$=20.42 $\mu$eV/atom, respectively. The other metastable structure is 50 meV/atom higher than that of *Pnma*.

The searched two lowest-energy $Co_3Si_2$ structures have negligible magnetism. All of the following stoichiometries are larger than 100 meV/atom in energy in comparison to their respective lowest-energy positions sited well at the convex-hull line, including $Co_8Si_3$ (~160), $Co_5Si_2$ (~100), $Co_7Si_3$ (~100), $Co_9Si_4$ (~150), $Co_9Si_5$ (~160), $Co_7Si_4$ (~160), $Co_8Si_5$ (~150), $Co_4Si_3$ (~100 meV/atom), in a word, these stoichiometries might have lower-energy structures at high pressure, thus it is not necessary to study these stoichiometries at ambient conditions as the currently available technology might be still unable to synthesize them. We show these lower-energy structures and MAE for reference purpose only. All of the low-energy $Co_8Si$ structures show small MAE below 20 $\mu$eV/atom and unperfect easy-axis/plane MAE. We ignore to discuss the $Co_9Si$ because its $E_f$ is about 0. The searched low-energy $Co_6Si$ structures indeed has large magnetism, whereas all of them haven't perfect easy-axis/plane MAE, only one structure shows nearly-perfect easy-axis MAE with values of $E_{001}$=19.89 and $E_{100}$=18.49 $\mu$eV/atom, respectively, which is about 30 meV/atom higher than that of convex-hull point, as is shown in Table S5.

$Co_5Si$ exists many large-magnetism low-energy structures, whereas none of them shows perfect easy-axis/plane MAE, the low/high-energy $Co_5Si$ structure is also dominated mainly by the low/high-energy Co $P6_3/mmc$ and $Fm\bar{3}m$ building blocks, respectively, that is, the energy order of the Co building block determines the energy order of the $Co_5Si$. We list several low-energy large-magnetism structures, *Ama*2 ($E_f$=-0.191, $E_{010}$=336.35, $E_{100}$=251.25), $Cmc2_1$ ($E_f$=-0.1646, $E_{001}$=321.31, $E_{010}$=9.34),

$P2$ ($E_f$=-0.1633, $E_{010}$=435.12, $E_{100}$=896.43), $I\bar{4}2d$ ($E_f$=-0.1374, $E_{001}$=180.58, $E_{100}$=532.35), *Cccm* ($E_f$=-0.1171 eV/atom, $E_{001}$=37.85 and $E_{010}$=191 $\mu$eV/atom), and so on, some of which could be synthesized due to the small energy difference between their own energies and the convex-hull value, as is shown in Table S5 and Figure 1.

There are several nearly-perfect easy-plane structures with energies about 70 meV/atom higher than that of the lowest-energy structure in $Co_5Si$, such as *Cc* ($E_f$=-0.1646, $E_{001}$=321.31, $E_{010}$=9.34). *Pmmm* ($E_f$=-0.1227, $E_{010}$=71.86, $E_{100}$=1.78). $P2$ ($E_f$=-0.146, $E_{001}$=127.83, $E_{010}$=7.5). *Cccm* ($E_f$=-0.1171, $E_{001}$=37.84, $E_{100}$=0.21), *Fddd* ($E_f$=-0.1626, $E_{001}$=90.31, $E_{010}$=5.5), *Amm2* ($E_f$=-0.1371 eV/atom, $E_{001}$=0.59, $E_{100}$=31.4373 $\mu$eV/atom), and so on, as is shown in Table S5. In addition, there are also several structures with nearly-perfect easy-axis such as $C2$, ($E_f$=-0.1395, $E_{001}$=66.63, $E_{100}$=71.5), $P2_1/c$ ($E_f$=-0.1492, $E_{010}$=20.81, $E_{100}$=17.79), $I4_1/amd$ ($E_f$=-0.1175, $E_{001}$=46.8, $E_{001}$=39.91). *Pbcm* ($E_f$=-0.1383 eV/atom, $E_{010}$=42.94, $E_{100}$=35.17 $\mu$eV/atom), and so on, as are shown in Tables S5,6. In a word, $Co_5Si$ is an important compound.

The searched two lowest-energy $Co_9Si_2$ *Pbcm* ($E_f$=-0.1272) and $P1$ ($E_f$=-0.1275 eV/atom) are about 100 meV/atom higher than the convex-hull value, as is shown in Table S6, thus we ignore to analysize them. Several $Co_4Si$ have similar $E_f$ (-0.2 eV/atom), which are about 50 meV/atom higher than convex-hull value, including $R\bar{3}$ ($E_{001}$=73.12, $E_{100}$=64.37), $P2_13$ ($E_{001}$=75.19, $E_{010}$=380.39), $I4/m$ ($E_{010}$=60.73, $E_{100}$=65.47), $I4/m$ ($E_{010}$=44.13, $E_{100}$=81.44 $\mu$eV/atom), in which one or two structures have nearly-perfect easy-axis MAE, as is shown in Table S5. Furthermore, $P4_1$ has a large MAE ($E_{010}$=220.85, $E_{100}$=190.99 $\mu$eV/atom) but it is about 100 meV/atom higher than convex-hull value with a $E_f$ of -0.1542 eV/atom, as is shown in Table S6. Their building blocks are also dominated by the largely distorted Co lattice.

Some $Co_7Si_2$ might be synthesized when comparing its energy with that of available metastable $Co_3Si$ ($P6_3/mmc$) and $Co_4Si$, as is shown in Table S5 and Figure 1, including a nearly-perfect easy-axis MAE $P\bar{1}$ ($E_f$=-0.2118, $E_{010}$=36.15, $E_{100}$=36.96), $P4_2/ncm$ ($E_f$=-0.2053, $E_{001}$=43.63, $E_{100}$=55.11) and $P2_12_12_1$ ($E_f$=-0.1911 eV/atom, $E_{001}$=2.19, $E_{100}$=50.21 $\mu$eV/atom), respectively.

The searched two lowest-energy $Co_5Si_3$ *Pbam* ($E_f$=-0.481) and *Cmmm* ($E_f$=-0.4737 eV/atom) are about 2~3 meV/atom higher than that of convex-hull value, whereas their MAE are negeligible. Both $Fe_5Si_3$ and $Mn_5Si_3$ $P6_3/mcm$ phase have been synthesized experimentally, the calculated $E_f$ is -0.4212 eV/atom in $Co_5Si_3$ $P6_3/mcm$, meaning a metastable phase. Our previous calculations show that the experimental metastable phase $Fe_5Si_3$ ($P6_3/mcm$) exhibits a large easy-plane magnetic anisotropy with a value of 157 $\mu$eV/atom ($E_{100}=E_{010}<E_{001}$). The experimentally synthesized nanoscale $Mn_5Si_3$ $P6_3/mcm$ shows interesting properties, we thus calculated its MAE and obtain the $E_{010}$=42.48 and $E_{100}$= 37.31 $\mu$eV/atom, respectively. The present $Co_5Si_3$ $P6_3/mcm$ phase shows $E_{001}$=15.11 and $E_{100}$=12.36 $\mu$eV/atom, respectively, as is shown in Table S1. Fe atoms have two different kinds of magnetic moment in $Fe_5Si_3$ ($P6_3/mcm$) with values of 1.833, 1.796, 1.838, 1.318, 1.313 $\mu_B$, the total magnetic moment is 15.756 $\mu_B$ in unit cell, as are also the cases in $Mn_5Si_3$ and $Co_5Si_3$, it is 0.722, 0.719, 0.669, 0.299, 0.258 $\mu_B$ in $Mn_5Si_3$ with a total value of 5.148 $\mu_B$, it is 0.815, 0.812, 0.81, 0.31, 0.281 $\mu_B$ in $Co_5Si_3$ with a total value of 5.967 $\mu_B$. This phenomenon clearly demonstrates that the magnetism shows strong lattice framework or the atomic coordinate dependence. $Co_5Si_3$ has larger magnetism than $Mn_5Si_3$. The atomic radius of Mn is slightly longer than that of Co with a value of about 5%[19-21], therefore the lattice volume of $Mn_5Si_3$ is about 2% larger than that of $Co_5Si_3$, previous study found that the shorter atomic distance corresponds to the smaller atomic moment in iron silicides[1]. The larger average moment contributes to the larger MAE is easily understandable in $Fe_5Si_3$. However, the larger average moment of Co in $Co_5Si_3$ relative to that of Mn in $Mn_5Si_3$ contributes to far smaller (13.73 $\mu$eV/atom) average MAE than the latter (39.90 $\mu$eV/atom) under the individual stable $P6_3/mcm$ lattice, which couldn't be uniquely illustrated by the atomic distance. To reveal the possible origin we use the same $P6_3/mcm$ lattice framework for $X_5Si_3$ (X=Mn, Fe, Co) to calculate and compare their MAE, as is shown in Table S2, clearly, the lattice shape and atomic coordinates crucially decide the MAE. For instance, the MAE is less than 20 $\mu$eV/atom in the relaxed $Co_5Si_3$ framework, which is also the case when the Co is substituted by Mn ($Mn_5Si_3$) and Fe ($Fe_5Si_3$), respectively. This is applicable to the relaxed $Mn_5Si_3$ (Mn $\rightarrow$ Co/Fe) and $Fe_5Si_3$ (Fe $\rightarrow$ Mn/Co). For any given framework and elemental species, the easy-axis direction is unchanged, meaning that the lattice shape determine the MAE. The lattice framework also determines the MAE value, such as the MAE in the relaxed $Co_5Si_3$ is always relatively small even for the substituted $Mn_5Si_3$ and $Fe_5Si_3$ when comparing to the case of $Fe_5Si_3$ as which always shows relatively large MAE whether substitution by Co/Mn or not, as are also the case for the relaxed $Mn_5Si_3$, still independent on the elemental species. This conclusion helps us to deeply understand the giant MAE of experimental nanoparticle $Co_3Si$, which might origin the largely distorted lattice shape and non-equilibrium atomic arrangements. Clearly, to obtain large MAE in nanoscale, the perfect bulk lattices must have large MAE. Only $Fe_5Si_3$ $P6_3/mcm$ in its stable status shows nearly-perfect easy-plane MAE ($E_{001}$=157.25, $E_{100}$=0.52 $\mu$eV/atom), as is shown in Table 4.

The lowest-energy $Co_7Si$ locates well at the convex hull profile, meaning these low-energy structures could by synthesized. For simplicity, we only list the structures with energy about 30 meV/atom lower than that of the lowest-energy one, in fact, too many structures site at an energy range of 30~100 meV/atom but these structures have non-perfect or nearly-perfect and large easy-axis/plane MAE, as is shown in Table S5. However, fortunately, two structures have nearly-perfect easy-axis MAE, one is $P4_132$ ($E_f$=-0.1188) with values of $E_{001}$=98.46 and $E_{010}$=92.28, the other one is $P\bar{1}$ ($E_f$=-0.1051 eV/atom) with values of $E_{001}$=37.86 and $E_{100}$=35.6 $\mu$eV/atom, respectively.

## G. $Mn_3Si$

For simplicity purpose, we ignore to plot the convex hull for the manganese silicides. Furthermore, we also ignore to plot the building block as the Mn is a very complicated crystal containing 58 atoms in the cubic cell in its ground state at ambient conditions. As is shown in table 5, the experimental $Fm\bar{3}m$-A ($E_f$=-0.3094) $Mn_3Si$ presents $E_{001}$=91.79 and $E_{010}$=61.78, respectively, the other experimental $P4/mmm$-A ($E_f$=-0.2103 eV/atom) $Mn_3Si$ presents $E_{010}$=72.02 and $E_{100}$=29.78 $\mu$eV/atom, respectively. The energy difference between $Fm\bar{3}m$-A and $P4/mmm$-A is about 100 meV/atom, indicating that the higher-energy metastable phase also could be synthesized. A structure ($Fm\bar{3}m$, $E_f$=-0.3096 eV/atom) with same MAE proportion with those of experimental ones along the three axes is searched with values of $E_{100}$=21.39 and $E_{001}$=14.25 $\mu$eV/atom, respectively. A perfect easy-axis MAE $Mn_3Si$ $P222_1$ ($E_f$=-0.2103 eV/atom) is searched with identical values of $E_{001}$=$E_{100}$=51.25 $\mu$eV/atom, a very promising magnetic material. Similarly, *Ibam* ($E_f$=-0.2207 eV/atom) also presents nearly-perfect MAE with individual values of $E_{001}$=25.98 and $E_{100}$=25.81 $\mu$eV/atom, a meaningful magnetic material. Two nearly-perfect easy-plane structures are searched, whose energies are nearly same with that of *P*4/*mmm*-A, including *Pmma* ($E_{001}$=0.01, $E_{100}$=51.69) and $P\bar{4}2_1m$ ($E_{010}$=1.15, $E_{100}$=20.58 $\mu$eV/atom). Most of the $Mn_3Si$ lattices are square shape, meaning the similar building block with that of $P4/mmm$-A. The other higher-energy and small-magnetism $Mn_3Si$ structures are shown in Table S7.

### H. $Mn_5Si_2$

As is shown in table 6, the experimental $Mn_5Si_2$ $P4_32_12$ presents ($E_f$=-0.2631 eV/atom) $E_{001}$=171.25 and $E_{010}$=23.93 $\mu$eV/atom, respectively. Two perfect easy-axis structures are searched, including *Ibam* ($E_f$=-0.261, $E_{010}$=$E_{100}$=19.21) and *I*4/*mmm* ($E_f$=-0.2014 eV/atom, $E_{001}$=$E_{100}$=30.13 $\mu$eV/atom), both structures are synthesizeable. Two perfect easy-plane MAE structures *Pbam* and *C*2/*c* are searched, which are about 100 and 80 meV/atom higher than that of $P4_32_12$ and show $E_{001}$=39.65 and $E_{100}$=6.25 $\mu$eV/atom, respectively. One structure $Cmc2_1$ with giant MAE is searched ($E_f$=-0.1777 eV/atom, $E_{001}$=728.63 and $E_{100}$=696.157 $\mu$eV/atom), which might also has some special applications despite it is not totally-perfect easy-axis MAE. In a word, $Mn_5Si_2$ is an important compound. The other searched $Mn_5Si_2$ structures are shown in Table S8.

**I. $Mn_5Si_3$**

As is shown in table 7, the experimental $Mn_5Si_3$ $P6_3/mcm$ ($E_f$=-0.3306 eV/atom) shows nearly-perfect easy-axis MAE, with values of $E_{010}$=49.11 and $E_{100}$=55.1 $\mu$eV/atom, respectively. A structure $I4_122$ ($E_f$=-0.6676 eV/atom) with energy about 300 meV/atom lower than that of experimental $P6_3/mcm$ is searched, with axial energies of $E_{001}$=8.13 and $E_{100}$=63.56 $\mu$eV/atom, respectively. A slight distortion increases MAE to $E_{001}$=121.64 and $E_{010}$=17.82 $\mu$eV/atom with a changed symmetry $P4_122$ but almost unchanged energy ($E_f$=-0.6675 eV/atom).

Two perfect easy-plane MAE *Pmma* and $P4_122$ with energies about 30 meV/atom higher than that of $P6_3/mcm$ are searched, with respective hard-axis energies of $E_{100}$=48.88 and $E_{001}$=65.58 $\mu$eV/atom. Four nearly-perfect easy-plane MAE structures with energies about 10~30 meV/atom higher than that of $P6_3/mcm$ are searched, including *I*4/*mmm* ($E_{010}$=0.40, $E_{100}$=20.55), *Pmmm* ($E_{010}$=12.48, $E_{100}$=0.81), *P*4/*mbm* ($E_{001}$=62.51, $E_{010}$=2.65), *Amm*2 ($E_{001}$=60.55, $E_{100}$=1.4 $\mu$eV/atom), respectively.

*Pbcn* ($E_f$=-0.3112, $E_{010}$=39.30, $E_{100}$=39.24) and *Pm* ($E_f$=-0.2524 eV/atom,

$E_{001}$=82.47, $E_{010}$=81.98 $\mu$eV/atom) present nearly-perfect easy-axis MAE and could be synthesized due to their lower energies. Therefore $Mn_5Si_3$ is an interesting compound. For simplicity, the other $Mn_5Si_3$ structures are shown in Table S9.

### J. $Mn_6Si$

Experimental $Mn_6Si$ structure has 53 atoms per unit cell ($Mn_{85.5}Si_{14.5}$), whereas it is 56 atoms in 8-unit cell, therefore we ignore to compare the $E_f$. All of the searched low-energy structures shown in Figure 1 haven't the experimental symmetry $R\bar{3}$, we thus systematically analyzed the cell shape and do find plenty of cells have the hexagonal shape, whereas we didn't find the experimentally reported rhombohedral shape cell, in fact lots of the searched cell are square shape, therefore we didn't deeply study the experimentally reported structure. The searched nearly-perfect easy-plane *Pbcn* and *Pbca* have nearly same energies, with a $E_f$ of -0.137 eV/atom and the hard-axis energy of about 54 $\mu$eV/atom, as is shown in Table 8, in which the hard-axis orientation could be switched due to the structural distortion between them, in fact, this is a frequently occurred phenomenon. The next lowest-energy $P2_12_12_1$ ($E_f$=-0.099 eV/atom) presents $E_{010}$=25.03 and $E_{100}$=76.82 $\mu$eV/atom, respectively. The even higher-energy $Pna2_1$ ($E_f$=-0.0877 eV/atom) shows larger MAE with values of $E_{010}$=241.51 and $E_{100}$=96.65 $\mu$eV/atom, respectively. None perfect easy-axis MAE structure is found in $Mn_6Si$, several nearly-perfect easy-plane MAE are found but their energies are about 50 meV/atom higher than that of the lowest-energy structure, thus we only discuss several low-energy structures.

### K. $Mn_9Si_2$

We also ignore to compare the experimental $Mn_9Si_2$ ($Mn_{81.5}Si_{18.5}$) as it has 186 atoms in the unit cell, whereas it has 176 atoms in 16-unit cell. As is shown in Figure 1, the low-energy $Mn_9Si_2$ might be synthesized if its decomposed products are metastable phases of $Mn_4Si$ and $Mn_5Si$, thus we list three low-energy structures, as is shown in Table 8, clearly, all of them show medium or small non-perfect easy-axis/plane MAE, such as $P4_12_12$ has lowest energy with axial energies of $E_{001}$=143.17 and $E_{010}$=94 $\mu$eV/atom, respectively.

**L. MAE of $Mn_3Ge$, $Mn_5Ge_2$, $Mn_5Ge_3$, $Mn_6Ge$, $Mn_9Ge_2$ calculated from their respective perfect Si-containing counterparts**

The perfect easy-axis/plane MAE in Si-containing compounds are calculated for Ge-containing counterparts, unfortunately, only several structures have negative $E_f$, thus we didn't carefully discuss them as all of the calculations are done by direct substitution in the coordinate file. The experimental $Mn_3Si$ $Fm\bar{3}m$ presents $E_{001}$=91.42 and $E_{010}$=61.88 $\mu$eV/atom, but it is $E_{010}$=25.55 and $E_{100}$=96.15 $\mu$eV/atom in $Mn_3Ge$ ($E_f$=-0.0971 eV/atom). Similarly, $E_{010}$=69.84 and $E_{100}$=137.2 $\mu$eV/atom in $Mn_3Si$ corresponds to $E_{010}$=214.93 and $E_{100}$=207.81 $\mu$eV/atom in $Mn_3Ge$ ($E_f$=-0.0542 eV/atom), displaying a nearly-perfect and very large easy-axis MAE. In addition, $E_{010}$=39.3 and $E_{100}$=39.24 $\mu$eV/atom in $Mn_5Si_3$ corresponds to 25.98 and 26.22 $\mu$eV/atom in $Mn_5Ge_3$ ($E_f$=-0.0885 eV/atom), also presenting a nearly-perfect easy-axis MAE. Similarly, $E_{001}$=121.64 and $E_{010}$=17.82 $\mu$eV/atom in $Mn_5Si_3$ corresponds to 96.25 and 58.12 in $Mn_5Ge_3$ ($E_f$ =-0.5409 eV/atom). Meanwhile, $E_{001}$=143.17 and $E_{010}$=94 $\mu$eV/atom in $Mn_9Si_2$ corresponds to 166.81 and 135.68 in $Mn_9Ge_2$ ($E_f$=-0.0121 eV/atom). In a word, Ge-containing compounds also have several structures with ideal MAE.

## IV. Conclusions

Our calculated bulk lattice failed to repeat the previously calculated and measured giant magnetic properties of the nanoscale $Co_3Si$ $P6_3/mmc$ even for the lattice parameter. Five lower-energy structures than that of metastable $Co_3Si$ $P6_3/mmc$ are searched, in which $I4/mmm$ presents perfect easy-plane magnetocrystalline anisotropy energy (MAE) with a hard-axis energy of 221.61 $\mu$eV/atom, the lowest-energy *Cmcm* shows nearly-perfect easy-plane MAE with a medium energy of $E_{010}$=0.24 and $E_{100}$=28.46 $\mu$eV/atom in comparison with that of the easy-axis $E_{001}$, another *Cmcm* shows non-perfect large MAE with axial energies of $E_{010}$=150.27 and $E_{100}$=6.36 $\mu$eV/atom, *Pmmn* and $P2_1$ show non-perfect large MAE between 100~200 $\mu$eV/atom, $I4/mcm$ shows nearly-perfect easy-plane MAE with values of $E_{001}$=2.81 and $E_{100}$= 211.67 $\mu$eV/atom in comparison with that of the easy-axis $E_{010}$. A structure $Pm\bar{3}m$ with similar energy with that of $P6_3/mmc$ also shows nearly-perfect easy-axis MAE with a value of about 70 $\mu$eV/atom.

All of the $Co_3C$ and $Co_3Sn$ present positive formation of energy. The MAE between $Co_3Si$ and $Co_3Ge$ are similar in the lowest-energy *Cmcm*. $Co_3Ge$ has similar symmetry with that of $Co_3Si$, two nearly-perfect easy-axis MAE structures with values within 110~120 $\mu$eV/atom are searched with energies about 40 meV/atom higher than that of *Cmcm*, three nearly-perfect easy-axis MAE structures with values of ~25 $\mu$eV/atom are searched with energies of 60~70 meV/atom higher than that of *Cmcm*, a perfect easy-plane MAE *I*4/*mmm* with values of ~157 $\mu$eV/atom is searched with energy of ~30 meV/atom higher than that of *Cmcm.* Four nearly-perfect easy-plane MAE structures with energies of 60~70 meV/atom higher than that of *Cmcm* are searched and show diverse MAE values. $Co_3Ge$ doesn't present easy-axis MAE in *P*6$_3$/*mmc*. $Co_3Ge$ $P\bar{1}$ presents nearly-perfect easy-axis MAE, which is different with that of $Co_3Si$.

A low-energy Co $R\bar{3}m$ is searched, with a lower energy than the experimentally metastable $Fm\bar{3}m$, which presents nearly-perfect easy-axis MAE. Many low-energy $Co_5Si$ have hundreds of MAE but none of them shows perfect MAE except for several nearly-perfect easy-axis/plane MAE structures. The several low-energy Co lattice building blocks generally control the energy order of the searched individual cobalt silicides, namely, the Co crystal framework keeps almost unchanged even after some substitutions by the other elements. Our comparison computation confirmed that the MAE shows strong lattice framework or the atomic coordinate dependence, as is further confirmed by the searched structures with similar energy but without similar MAE. Many $Mn_3Si$, $Mn_5Si_2$, $Mn_5Si_3$ exist perfect or nearly-perfect low-energy easy-axis/plane structures, as is also the case in the several Ge-containing counterparts.

Four nearly-perfect easy-axis/plane MAE with medium values are searched in $Mn_3Si$. Two perfect easy-axis MAE with medium value is searched in $Mn_5Si_2$, two nearly-perfect easy-plane MAE with medium values are searched in $Mn_5Si_2$, one $Mn_5Si_2$ structure with nearly-perfect easy-axis and giant MAE values (about 700 $\mu$eV/atom) is searched. A lower-energy structure with energy about 300 meV/atom lower than that of experimental $Mn_5Si_3$ is searched, together with three or four nearly-perfect easy-axis MAE structures and two perfect easy-plane and four nearly-perfect easy-plane MAE structures. All of the above mentioned structures are about 100 meV/atom lower than that of the searched lowest-energy structure or experimental structure. In a word, most of the currently searched easy-axis MAE structures could be synthesized experimentally, these easy-axis structures have very promising application in the diverse requirements field of spin-related magnetic device.

**Data and materials availability:**

The data that supports the findings of this study are available within the article and its supplementary material.

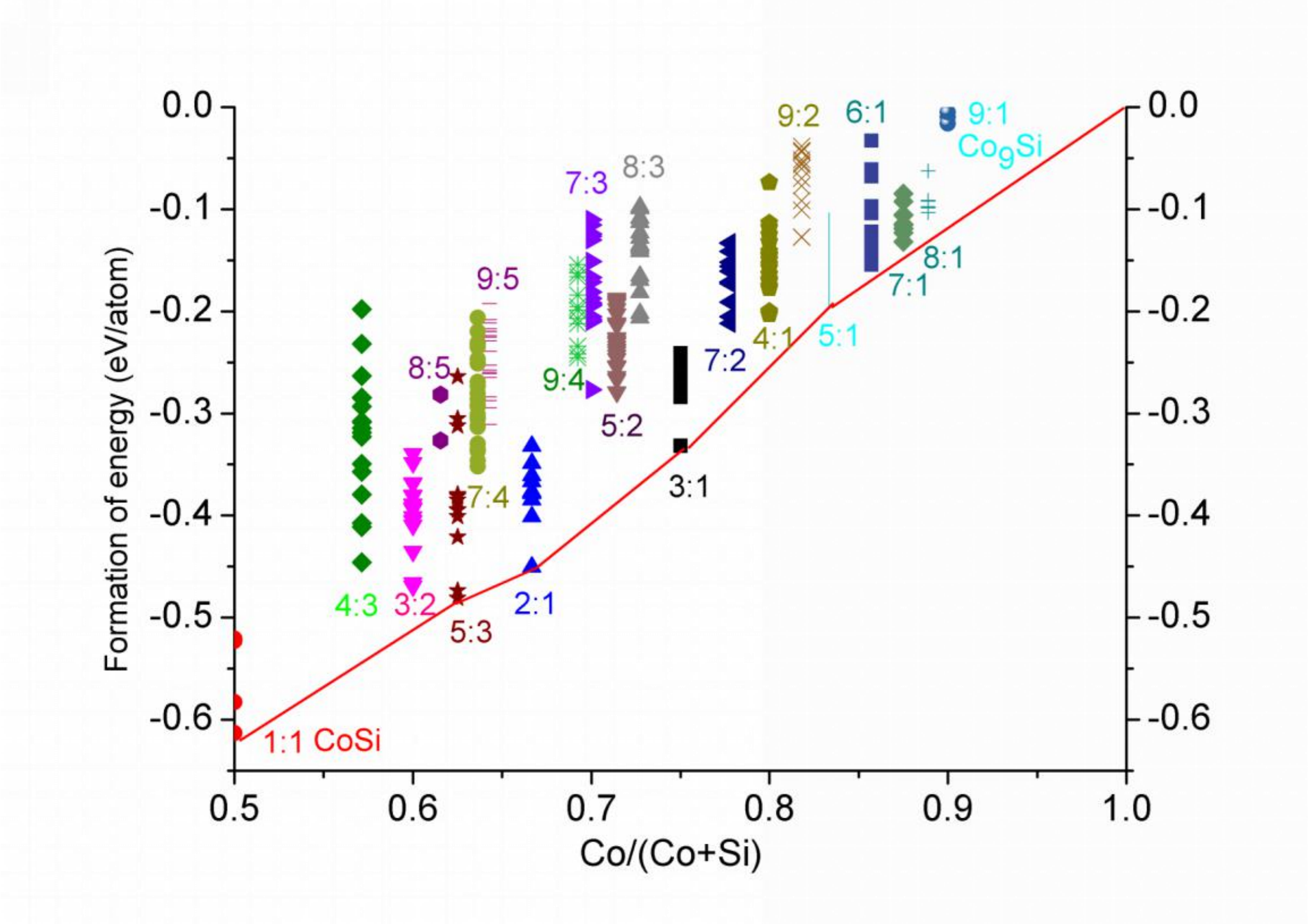

Figure 1, the formation of energy ($E_f$) of the binary cobalt silicides, in which we ignored the experimental data for clear purpose, including $Co_3Si$ ($E_f$=-0.2638 eV/atom, $P6_3/mmc$), $Fe_5Si_3$ ($E_f$=-0.3787 eV/atom, $P6_3/mcm$), respectively.

Table 1, the lattice parameter (Å), angle (˚), symmetry, atomic average energy $\overline{E}$ (eV/atom), building block, MAE ($\mu$eV/atom) of $Co_3Si$. All of the colors online only.

| Symm.<br>$\overline{E}$ | building block | $a$<br>$b$<br>$c$ | $\alpha$<br>$\beta$<br>$\gamma$ | Unit cell<br>Co Si | $E_{001}$<br>$E_{010}$<br>$E_{100}$ |
|---|---|---|---|---|---|
| $P6_3/mmc$<br>-6.9<br>Exp. | 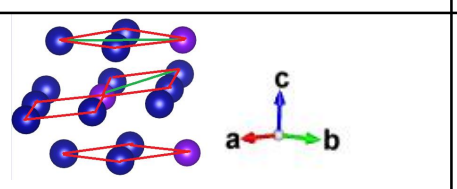 | 4.97<br>4.97<br>3.97 | 90<br>90<br>120 | 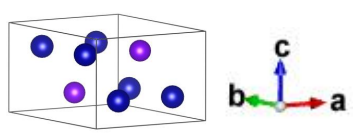 | 261.14<br>261.14<br>0 |
| $I4/mmm$-A<br>-6.9197 | 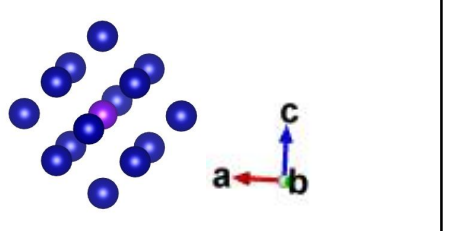 | 5.05<br>6.66<br>5.05 | 90<br>90<br>90 | 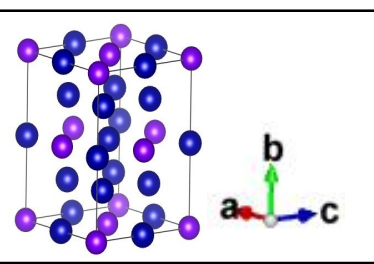 | 0<br>141.42<br>0 |
| $I4/mmm$-B<br>-6.9181 | 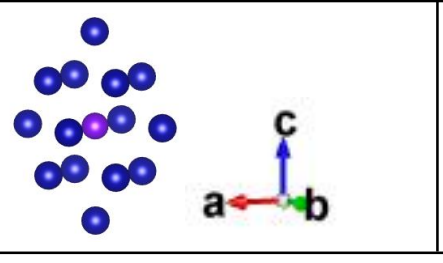 | 3.57<br>3.57<br>6.66 | 90<br>90<br>90 | 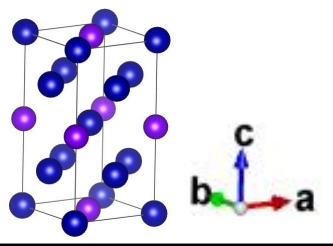 | 0<br>0<br>221.61 |
| *I*4/*mmm*-C<br>-6.9197 | 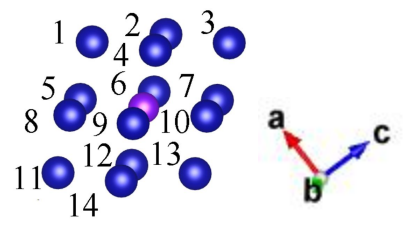 | 5.06<br>5.06<br>13.29 | 90<br>90<br>90 | 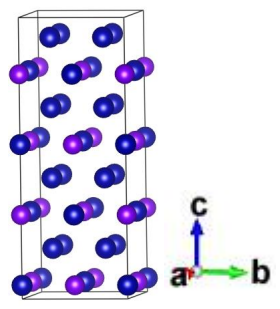 | 153.08<br>25.22<br>0 |
| *I*4/*mmm*-D<br>-6.9197 | 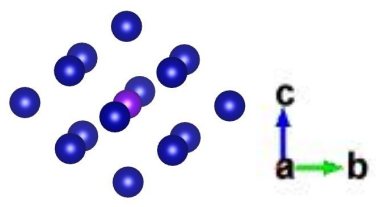 | 5.05<br>6.65<br>5.05 | 90<br>90<br>90 | 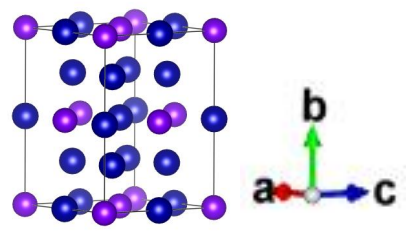 | 0.03<br>148.42<br>0 |
| *I*4/*mmm*-E<br>-6.8823 | 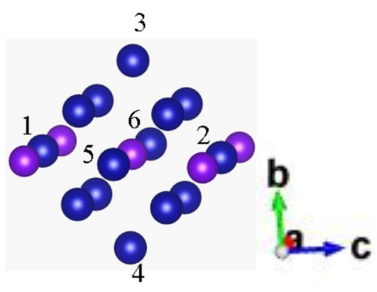 | 5.61<br>5.61<br>10.98 | 90<br>90<br>90 | 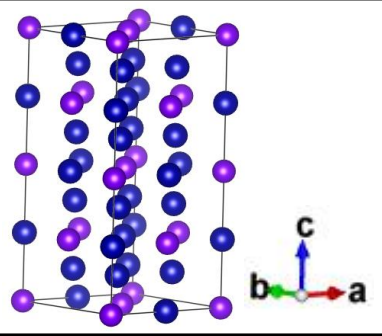 | 46.9<br>0<br>114.09 |
| *I*4/*mmm*-F<br>-6.8822 | 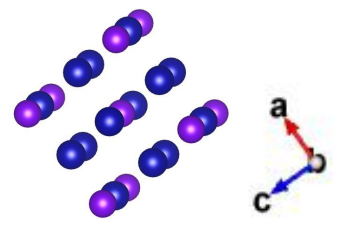 | 7.92<br>11.01<br>7.93 | 90<br>90<br>90 | 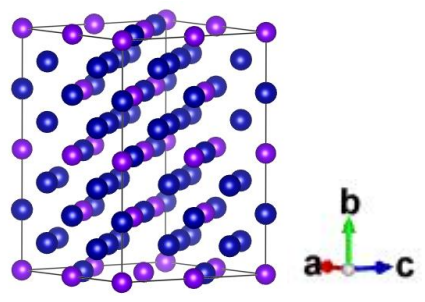 | 0<br>2.67<br>0.06 |
| *Cmcm*-A<br>-6.9673 | 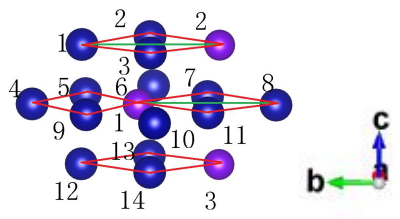 | 6.26<br>7.41<br>3.68 | 90<br>90<br>90 | 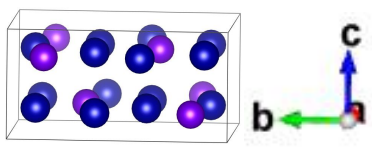 | 0<br>0.24<br>28.46 |
| *Cmcm*-B<br>-6.9183 | 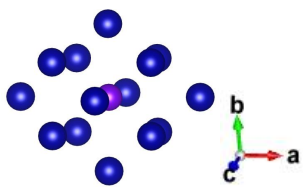 | 6.36<br>5.13<br>5.21 | 90<br>90<br>90 | 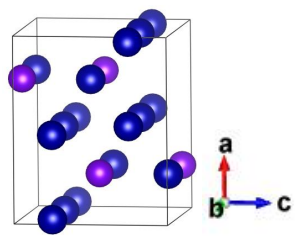 | 0<br>150.27<br>6.36 |
| *Cmcm*-C<br>-6.8863 | 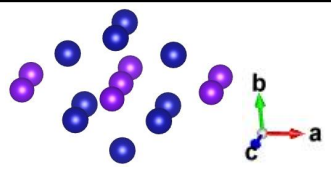 | 7.21<br>9.31<br>2.57 | 90<br>90<br>90 | 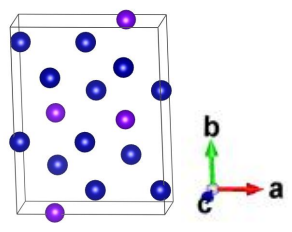 | 76.81<br>67.12<br>0 |
| *Cmcm*-D<br>-6.8815 | 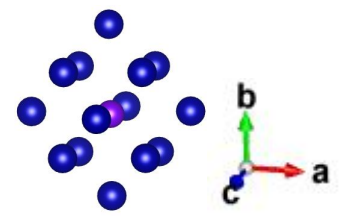 | 5.49<br>5.49<br>5.49 | 90<br>90<br>90 | 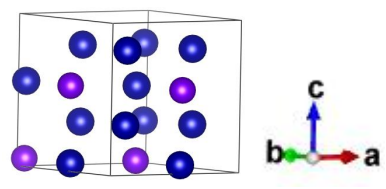 | 0<br>97.85<br>198.27 |
| *Pmmn*<br>-6.9143 | 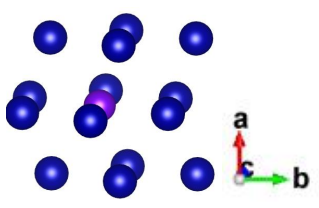 | 4.01<br>4.27<br>4.97 | 90<br>90<br>90 | 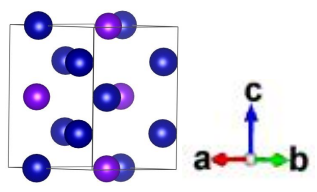 | 0<br>75.76<br>124.73 |
| *I*4/*mcm*<br>-6.9025 | 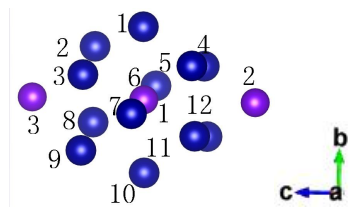 | 4.84<br>4.84<br>7.32 | 90<br>90<br>90 | 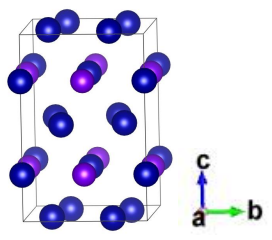 | 2.81<br>0<br>211.67 |
| $P2_1$<br>-6.9071 | 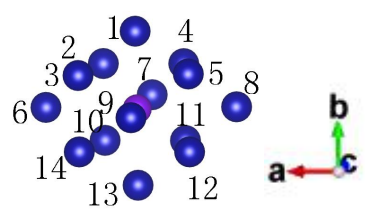 | 6.21<br>5.02<br>5.43 | 90<br>90.69<br>90 | 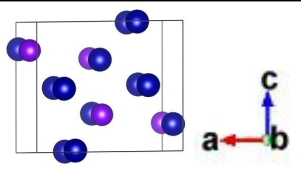 | 81.08<br>0<br>137.16 |
| $P6_3$/*mmc*-B<br>-6.8931 | 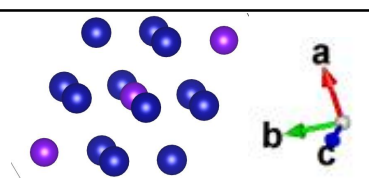 | 4.96<br>8.59<br>8.01 | 90<br>90<br>90 | 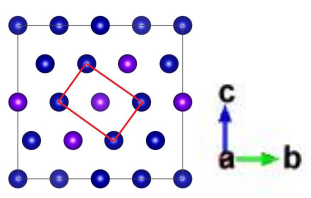 | 80.28<br>0<br>0.06 |
| *P*4/*mbm*-A<br>-6.8976 | 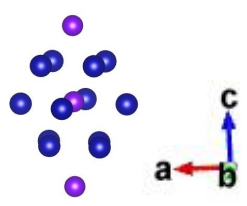 | 4.83<br>4.83<br>7.21 | 90<br>90<br>90 | 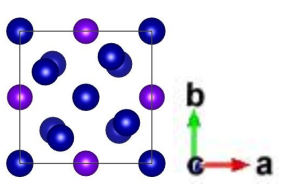 | 2.14<br>0<br>144.51 |
| $Pm\bar{3}m$ -A<br>-6.8891 | 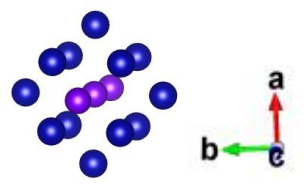 | 4.91<br>9.83<br>3.47 | 90<br>90<br>90 | 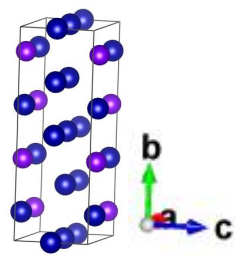 | 0<br>71.9<br>72.1 |
| $Pm\bar{3}m$ -B<br>-6.8876 | 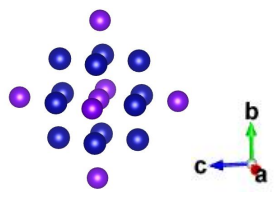 | 3.47<br>3.47<br>3.47 | 90<br>90<br>90 | 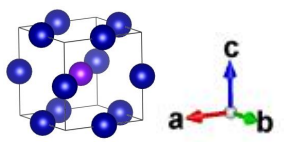 | 0<br>0<br>32.14 |

Table 2, the lattice parameter (Å), angle (˚), symmetry, formation of energy $E_f$ (eV/atom), MAE ($\mu$eV/atom) of $Co_3Ge$.

| Symm.<br>$E_f$ | $a$<br>$b$<br>$c$ | $\alpha$<br>$\beta$<br>$\gamma$ | Co Ge 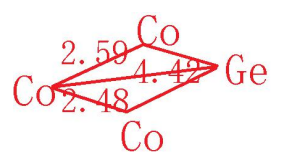 | $E_{001}$<br>$E_{010}$<br>$E_{100}$ |
|---|---|---|---|---|
| *Cmcm*<br>-0.1137 | 7.52<br>6.44<br>3.76 | 90<br>90<br>90 | 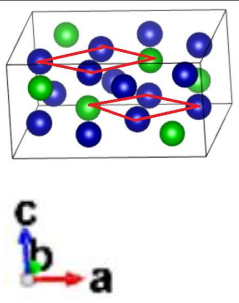 | 0<br>14.91<br>32.19 |
| *C*2/*m*,<br>-0.0603 | 8.62<br>5.12<br>6.22 | 90<br>138.35<br>90 | 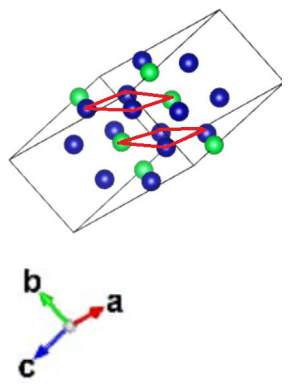 | 0<br>42.56<br>108.01 |
| *I*4/*mmm*<br>-0.0795 | 5.13<br>5.13<br>6.87 | 90<br>90<br>90 | 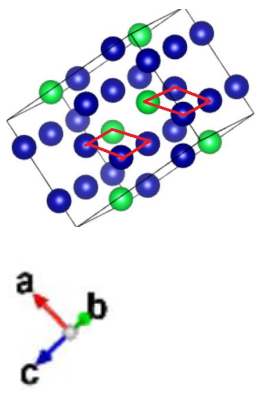 | 157.06<br>0<br>0 |
| *I*4/*mmm*<br>-0.0524 | 5.11<br>13.85<br>5.11 | 90<br>90<br>90 | 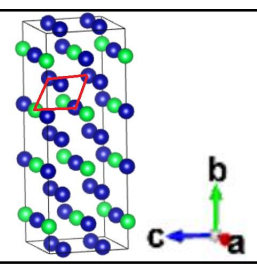 | 0<br>9.62<br>0 |
| *Pmmm*<br>-0.0728 | 5.97<br>5.09<br>5.97 | 89.99<br>93.2<br>90 | 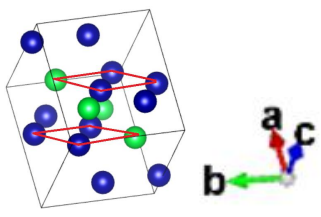 | 122.41<br>0<br>122.58 |
| $P\bar{1}$<br>-0.0694 | 4.09<br>4.35<br>5.08 | 90<br>89.98<br>90.03 | 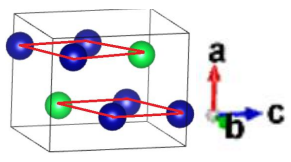 | 0<br>116.38<br>114 |
| *Cmmm*,<br>-0.0422 | 5.06<br>5.06<br>3.54 | 90<br>90<br>90 | 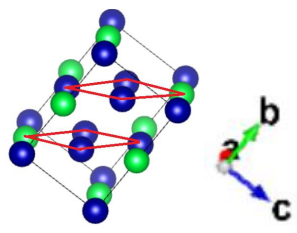 | 0<br>25.5<br>26.3 |
| $P2_12_12_1$,<br>-0.0337 | 5.69<br>5.69<br>5.66 | 90<br>90<br>90 | 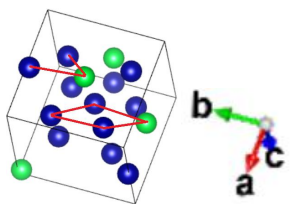 | 0<br>24.15<br>23.25 |
| *P*6₃/*mmc*,<br>-0.0542 | 5.07<br>5.07<br>4.07 | 90<br>90<br>119.99 | 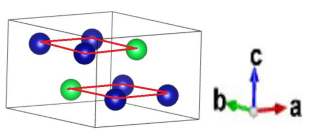 | 332.58<br>85.53<br>0 |
| *Cm*,<br>-0.0431 | 5.05<br>5.05<br>14.32 | 90<br>100.1<br>90 | 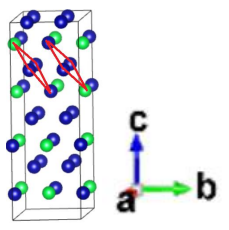 | 40.92<br>0<br>2.64 |
| *Ama*2,<br>-0.0532 | 8.79<br>5.07<br>8.14 | 90<br>90<br>90 | 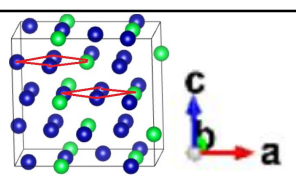 | 269.05<br>0<br>0.78 |
| *C*222₁,<br>-0.0513 | 8.76<br>5.07<br>8.18 | 90<br>90<br>90 | 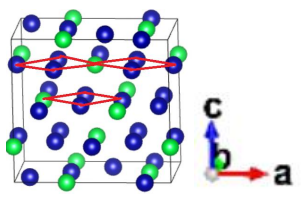 | 112.07<br>4.75<br>0 |
| *C*2/*m*,<br>-0.0472 | 7.16<br>7.10<br>5.05 | 90<br>135.01<br>90 | 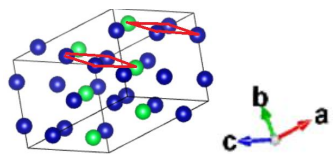 | 26.18<br>0<br>23 |

Table 3, the lattice parameter (Å), angle (˚), symmetry, atomic average energy $\overline{E}$ (eV/atom), building block or motif, MAE ($\mu$eV/atom) of Co.

| Sym.<br>$\overline{E}$ | motif | $a$<br>$b$<br>$c$ | $\alpha$<br>$\beta$<br>$\gamma$ | Unit cell | $E_{001}$<br>$E_{010}$<br>$E_{100}$ |
|---|---|---|---|---|---|
| $P6_3/mmc$<br>-7.0398 | 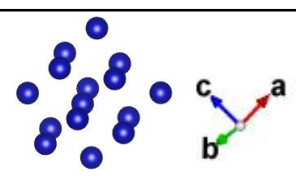 | 2.49<br>2.49<br>4.02 | 90<br>90<br>120 | 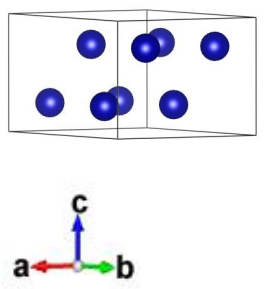 | 65.55<br>115.46<br>0 |
| $Fm\overline{3}m$<br>-7.0206 | 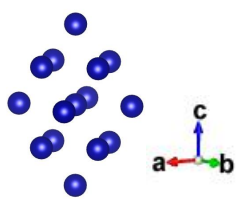 | 3.51<br>3.51<br>3.51 | 90<br>90<br>90 | 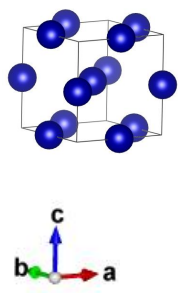 | 250.41<br>55.16<br>0 |
| $R\overline{3}m$<br>-7.0306 | 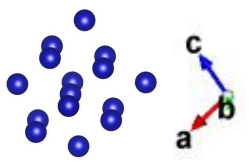 | 4.31<br>2.48<br>8.20 | 90<br>100.1<br>90 | 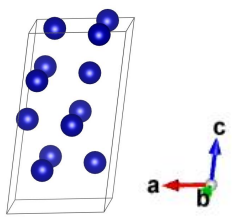 | 0<br>25.61<br>23.82 |
| $P4_2/mnm$<br>-6.9968 | 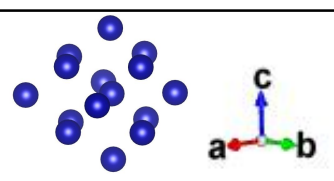 | 4.1<br>4.1<br>2.59 | 90<br>90<br>90 | 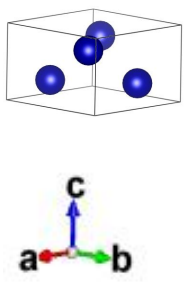 | 0<br>48.91<br>101.7 |
| *Cmca*<br>-6.9972 | 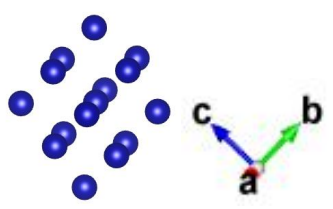 | 2.56<br>4.2<br>8.08 | 90<br>90<br>90 | 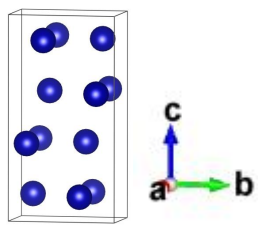 | 30.93<br>67.03<br>0 |
| Diamond Si<br>-5.4236 | 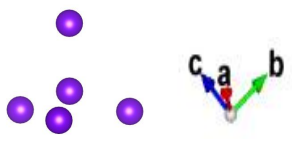 | 5.43<br>5.43 | 90<br>90 | 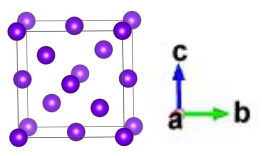 | |
| | | 5.43 | 90 | | |

Table 4, the searched lower-energy structure of $Fe_5Si_3$. Average atomic energy $\overline{E}$ (eV/atom), formation of energy $E_f$ (eV/atom), MAE ($\mu$eV/atom) and structure parameters (Å) and angles (°) are shown.

| Sym.<br>$\overline{E}$<br>$E_f$ | Fe<br>Si | $a$<br>$b$<br>$c$ | $\alpha$<br>$\beta$<br>$\gamma$ | $E_{001}$<br>$E_{010}$<br>$E_{100}$ |
|---|---|---|---|---|
| Exp.<br>$P6_3/mcm$<br>-7.5687<br>-0.3787 | 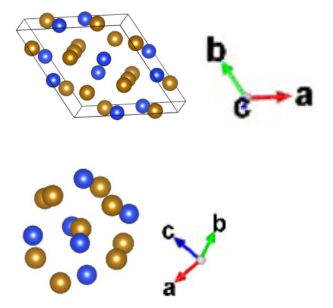 | 6.69<br>6.69<br>4.67 | 90<br>90<br>120 | 157.25<br>0<br>0.52 |
| Searched<br>*Cmcm*<br>-7.5973<br>-0.41 | 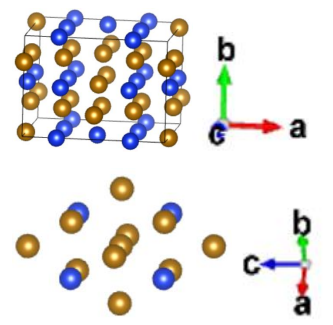 | 7.86<br>5.54<br>7.82 | 90<br>90<br>90 | 0.87<br>0<br>11.82 |
| *Pnma*,<br>-7.5916<br>-0.4044 | 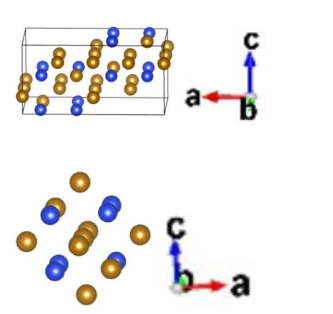 | 11.01<br>5.6<br>5.45 | 90<br>90<br>90 | 18.13<br>0<br>10.07 |
| $P2/m$,<br>-7.5580<br>-0.3708 | 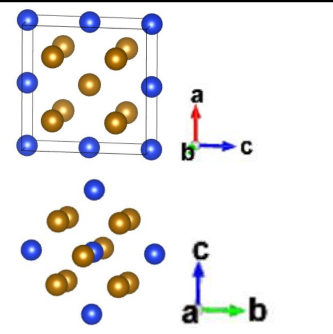 | 5.54<br>2.78<br>5.54 | 90<br>90<br>90 | 0.076<br>4.91<br>0 |
| *Fmmm*,<br>-7.5941<br>-0.4069 | 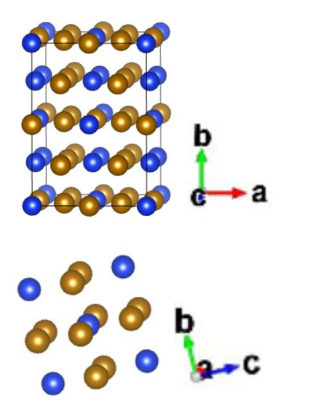 | 5.53<br>7.83<br>7.83 | 90<br>90<br>90 | 3.12<br>3.18<br>0 |
| $P\bar{1}$,<br>-7.4978<br>-0.3106 | 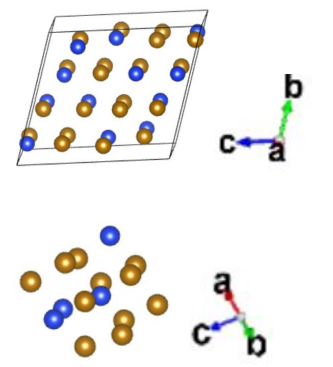 | 4.8<br>8.27<br>9.37 | 106.16<br>91.6<br>101.04 | 27.91<br>5.21<br>0 |
| *Cmca*,<br>-7.4763 | 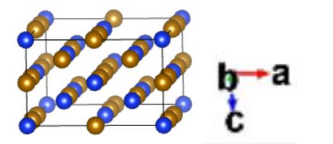 | 7.64<br>9.65 | 90<br>90 | 37.21<br>0 |
| -0.2891 | 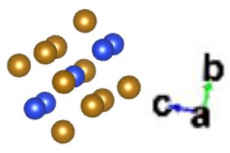 | 4.78 | 90 | 9.23 |
| $P2_12_12_1$,<br>-7.52683<br>-0.3395 | 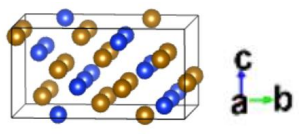<br>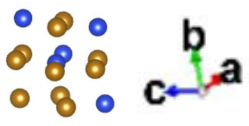 | 7.66<br>9.33<br>4.79 | 90<br>90<br>90 | 64.4<br>29.28<br>0 |

Table 5, the searched lower-energy structure of $Mn_3Si$. Formation of energy $E_f$ (eV/atom), MAE ($\mu$eV/atom) and structure parameters (Å) and angles (˚) are shown.

| | Sym.<br>$E_f$ | Unit cell<br>Mn<br>Si | $a$<br>$b$<br>$c$ | $\alpha$<br>$\beta$<br>$\gamma$ | $E_{001}$<br>$E_{010}$<br>$E_{100}$ |
|---|---|---|---|---|---|
| 1 | Exp.<br>$Fm\bar{3}m$,<br>-0.3094 | 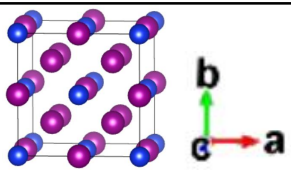 | 5.65<br>5.65<br>5.65 | 90<br>90<br>90 | 91.79<br>61.78<br>0 |
| 2 | Exp.<br>$P4/mmm$,<br>-0.2103 | 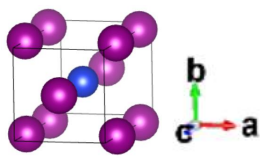 | 2.79<br>2.79<br>5.68 | 90<br>90<br>90 | 0<br>72.02<br>29.78 |
| 3 | searched<br>$P222_1$,<br>-0.2103 | 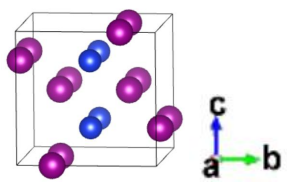 | 5.48<br>5.7<br>5.48 | 90<br>90<br>90 | 51.25<br>0<br>51.25 |
| 4 | $Ibam$,<br>-0.2207 | 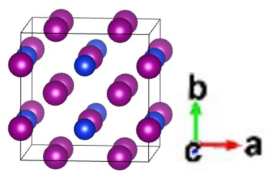 | 5.57<br>5.55<br>5.56 | 90<br>90<br>90 | 25.98<br>0<br>25.81 |
| 5 | $P\bar{4}2_1m$,<br>-0.2161 | 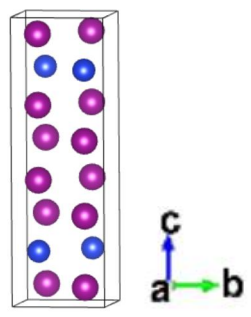 | 3.81<br>3.81<br>11.82 | 90<br>90<br>90 | 0<br>1.15<br>20.56 |

Table 6, the searched lower-energy structure of $Mn_5Si_2$. Formation of energy $E_f$ (eV/atom), MAE ($\mu$eV/atom) and structure parameters (Å) and angles (˚) are shown.

| | Sym.<br>$E_f$ | Unit cell<br>Mn<br>Si | $a$<br>$b$<br>$c$ | $\alpha$<br>$\beta$<br>$\gamma$ | $E_{001}$<br>$E_{010}$<br>$E_{100}$ |
|---|---|---|---|---|---|
| 1 | Exp.<br>$P4_32_12$<br>-0.2631 | 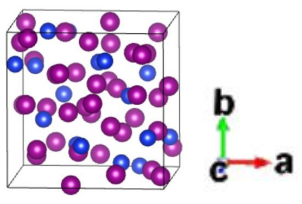 | 8.72<br>8.72<br>8.53 | 90<br>90<br>90 | 171.25<br>23.93<br>0 |
| 2 | searched<br>*Ibam*,<br>-0.2615 | 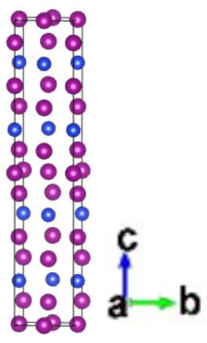 | 3.87<br>7.74<br>19.88 | 90<br>90<br>90 | 0<br>19.21<br>19.21 |
| 3 | *I*4/*mmm*,<br>-0.2014 | 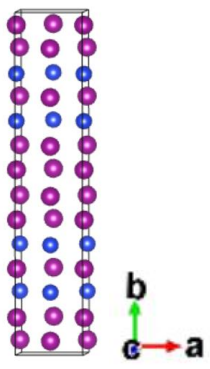 | 3.88<br>19.8<br>3.88 | 90<br>90<br>90 | 30.13<br>0<br>30.13 |
| 4 | *Pbam*,<br>-0.1678 | 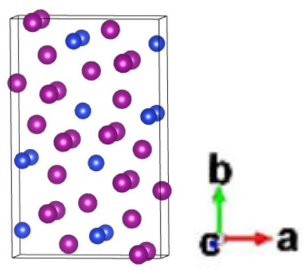 | 8.59<br>13.5<br>2.66 | 90<br>90<br>90 | 39.65<br>0<br>0 |
| 5 | *C*2/*c*<br>-0.1839 | 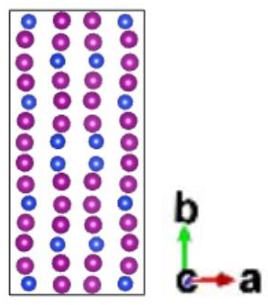 | 8.98<br>17.22<br>4.34 | 90<br>114.1<br>90 | 0<br>0<br>6.25 |
| 6 | $Cmc2_1$,<br>-0.1777 | 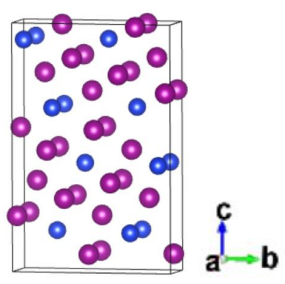 | 2.74<br>8.59<br>13.18 | 90<br>90<br>90 | 728.65<br>0<br>696.15 |

Table 7, the searched lower-energy structure of $Mn_5Si_3$. Formation of energy $E_f$ (eV/atom), MAE ($\mu$eV/atom) and structure parameters (Å) and angles (˚) are shown.

| | Sym.<br>$E_f$ | Unit cell<br>Mn Si | $a$<br>$b$<br>$c$ | $\alpha$<br>$\beta$<br>$\gamma$ | $E_{001}$<br>$E_{010}$<br>$E_{100}$ |
|---|---|---|---|---|---|
| 1 | Exp.<br>*P*6<sub></sub>3/*mcm*<br>-0.3306 | 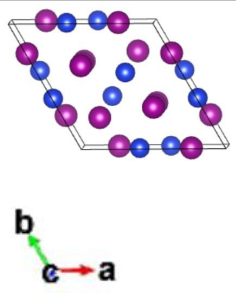 | 6.89<br>6.89<br>4.7889 | 90<br>90<br>120 | 0<br>49.11<br>55.1 |
| 2 | searched<br>*I*4<sub></sub>122,<br>-0.6676 | 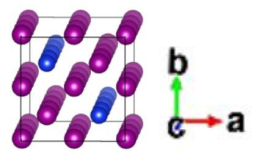 | 5.64<br>5.64<br>11.25 | 90<br>90<br>90 | 8.13<br>0<br>63.56 |
| 3 | *P*4122,<br>-0.6675 | 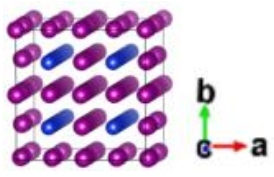 | 3.99<br>3.99<br>11.23 | 90<br>90<br>90 | 121.64<br>17.82<br>0 |
| 4 | *Pmma*,<br>-0.2986 | 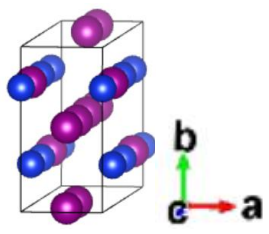 | 2.76<br>5.56<br>11.25 | 90<br>90<br>90 | 0<br>0<br>48.88 |
| 5 | *I*4/*mmm*,<br>-0.3194 | 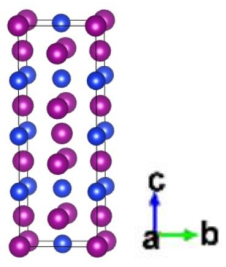 | 3.93<br>3.93<br>11.2 | 90<br>90<br>90 | 0<br>0.40<br>20.55 |
| 6 | *Pmmn*,<br>-0.3275 | 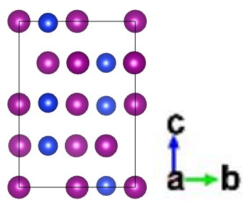 | 3.99<br>5.52<br>7.86 | 90<br>90<br>90 | 0<br>12.48<br>0.81 |
| 7 | *P*4122,<br>-0.2945 | 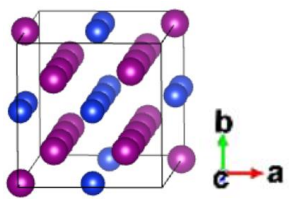 | 5.59<br>5.59<br>11.05 | 90<br>90<br>90 | 65.58<br>0<br>0 |
| 7 | *P*4/*mbm*,<br>-0.2941 | 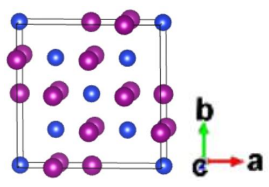 | 7.91<br>7.91<br>2.76 | 90<br>90<br>90 | 62.51<br>2.65<br>0 |
| 8 | *Amm*2,<br>-0.2943 | 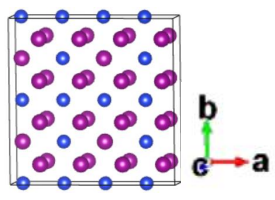 | 11.26<br>11.28<br>2.77 | 90<br>90<br>90 | 60.55<br>0<br>1.4 |
| 9 | *Pbcn*,<br>-0.3112 | 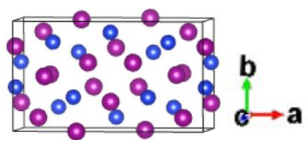 | 11.64<br>6.72<br>4.57 | 90<br>90<br>90 | 0<br>39.3<br>39.24 |
| 10 | *I*41/*amd*,<br>-0.3036 | 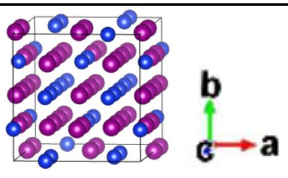 | 7.9<br>7.9<br>11.11 | 90<br>90<br>90 | 7.34<br>0<br>8.96 |
| 11 | *Pm*,<br>-0.2524 | 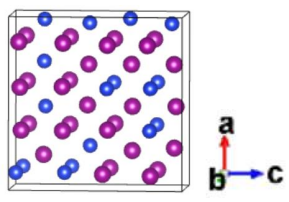 | 11.14<br>2.82<br>11.19 | 90<br>88.89<br>90 | 82.47<br>81.98<br>0 |

Table 8, the searched lower-energy structure of $Mn_6Si$ and $Mn_9Si_2$. Formation of energy $E_f$ (eV/atom), MAE ($\mu$eV/atom) and structure parameters (Å) and angles (˚) are shown.

| | Sym.<br>$E_f$<br>$Mn_6Si$ | Unit cell<br>Mn 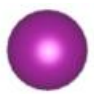 Si 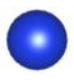 | $a$<br>$b$<br>$c$ | $\alpha$<br>$\beta$<br>$\gamma$ | $E_{001}$<br>$E_{010}$<br>$E_{100}$ |
|---|---|---|---|---|---|
| 1 | *Pbca*<br>-0.1371 | 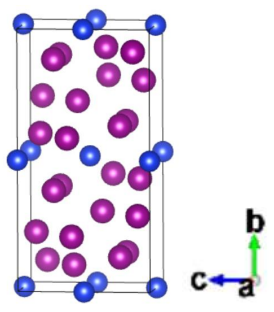 | 4.39<br>11.63<br>5.9 | 90<br>90<br>90 | 54.03<br>0.54<br>0 |
| 2 | *Pbcn*<br>-0.137 | 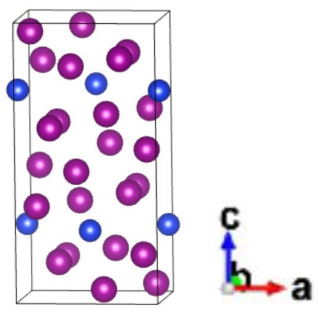 | 5.93<br>4.39<br>11.67 | 90<br>90<br>90 | 0.67<br>0<br>54.28 |
| 3 | $P2_12_12_1$<br>-0.0991 | 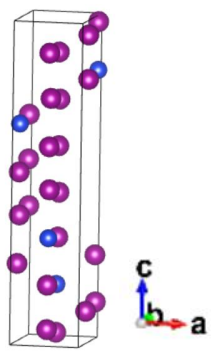 | 3.68<br>4.8<br>17.07 | 90<br>90<br>90 | 0<br>25.03<br>76.82 |
| 4 | $Pna2_1$<br>-0.0877 | 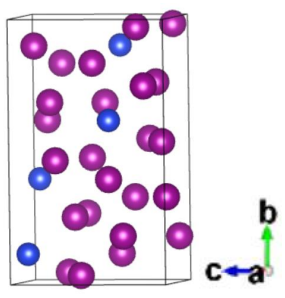 | 4.4<br>10.63<br>6.4 | 90<br>90<br>90 | 0<br>241.51<br>96.65 |
| | $Mn_9Si_2$ | | | | |
| 1 | $P4_12_12$<br>-0.1391 | 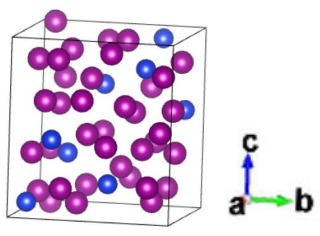 | 7.76<br>7.76<br>8.53 | 90<br>90<br>90 | 143.17<br>94<br>0 |
| | *Pbcn*<br>-0.1225 | 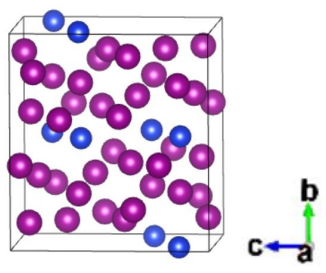 | 6.21<br>9.21<br>8.45 | 90<br>90<br>90 | 0<br>9.6<br>9.72 |
| 2 | $Pca2_1$<br>-0.1053 | 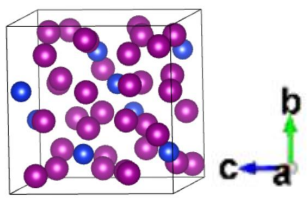 | 8.4<br>7.6<br>7.52 | 90<br>90<br>90 | 30.39<br>67.6<br>0 |